\documentclass[runningheads]{llncs}
\usepackage[T1]{fontenc}
\usepackage{graphicx}
\begin{document}
\title{Perceptions of AI-CBT: Trust and Barriers in Chinese Postgrads}
%
%
\author{Chan-in Sio\inst{1} \and
Alex Mann\inst{1} \and
Lingxi Fan\inst{1} \and
Andrew Cheung\inst{1}\and
Lik-hang Lee\inst{1*}}
\authorrunning{Chan-in SIO et al.}
%
\institute{The Hong Kong Polytechnic University, Hong Kong SAR, PRC\\
\email{*Corresponding author: lik-hang.lee@polyu.edu.hk}\\}
\maketitle              
\begin{abstract}
The mental well-being of graduate students is an increasing concern, yet the adoption of scalable support remains uneven. Artificial intelligence-powered cognitive behavioral therapy chatbots (AI-CBT) offer low barrier help, but little is known about how Chinese postgraduates perceive and use them. This qualitative study explored perceptions and experiences of AI-CBT chatbots among ten Chinese graduate students recruited through social media. Semi-structured Zoom interviews were conducted and analyzed using reflexive thematic analysis, with the Health Belief Model (HBM) and the Theory of Planned Behavior (TPB) as sensitizing frameworks. The findings indicate a cautious openness to AI-CBT chatbots: perceived usefulness and 24/7 access supported favorable attitudes, while data privacy, emotional safety, and uncertainty about `fit' for complex problems restricted the intention to use. Social norms (e.g., stigma and peer views) and perceived control (digital literacy, language quality) further shaped adoption. The study offers context-specific information to guide the culturally sensitive design, communication, and deployment of AI mental well-being tools for student populations in China and outlines the design implications around transparency, safeguards, and graduated care pathways.

\end{abstract}
\section{Introduction}\label{sec1}
Postgraduate student mental wellbeing has become a well-documented global concern, with consistently reported elevated rates of anxiety, depression, and suicidal ideation in this population \cite{CGSJed2024,Forrester2021,Woolston2017}. In China, these risks are compounded by intense academic competition, heavy workloads, and high stakes evaluation, while the formal seeker status of help remains stigmatized and comparatively low \cite{ning2024chinese,zhang2025mental}. Recent syntheses estimate the prevalence of anxiety and depression among Chinese university students at roughly 13.7\% and 20.8\%, respectively \cite{yang2024relationship}, with qualitative and survey evidence convergent from one-fifth to one-third experiencing notable distress \cite{ning2024chinese,bai2025mechanisms}. Longitudinal data further suggest a deterioration in symptoms during the course of training when supervisory support is limited \cite{wang2024relationship}. Despite the need, service utilization in major cities remains modest (e.g., $\sim$ 2.7----3. 1\%), leaving many students without professional support \cite{gearing2024mental,zhang2025mental}. These conditions create a practical and ethical imperative to explore scalable, lower-stigma forms of support.

Currently, rapid advances in artificial intelligence (AI) have expanded the scope of digital mental health beyond therapist-provided care to include screening, self-monitoring, stress management, and maintenance of daily well-being. AI-enabled systems promise on-demand access, personalization, and potential cost effectiveness. Within this space, AI-powered cognitive behavioral therapy (AI-CBT) chatbots have attracted particular attention as a self-help modality that can be private, always available and resource-efficient. Emerging evidence indicates that chatbot-enabled or AI-assisted interventions can reduce psychological distress and maintain user engagement \cite{He2022,mcfadyen2024ai}, in line with broader reviews of conversational agents and digital mental health \cite{Boucher2021,Higgins2023,Inkster2018}. However, adoption within specific high-need groups, such as Chinese graduate students, remains uncertain.

Two limitations in the current evidence motivate this study. First, much of what is known about attitudes towards AI in mental health is derived from clinical contexts or general population surveys \cite{Varghese2024,Benda2024}. These studies map broad patterns, interest in convenience and privacy, concerns about precision, safety and `loss of human touch', but provide limited information on how a particular student group, situated in a specific cultural and institutional environment, makes sense of and decides about AI-CBT in everyday life. Second, many attitudinal datasets predate recent advances in generative AI (e.g., more naturalistic dialogue and greater personalization), which may change user expectations and concerns \cite{Boucher2021,Higgins2023}. Therefore, a context-sensitive examination is warranted to understand how Chinese graduate students appraise AI-CBT chatbots and what facilitates or inhibits their uptake.

\paragraph{Aim and Research Question.}
This paper investigates how Chinese graduate students perceive and experience AI-CBT chatbots for mental well-being, with particular attention to trust, perceived effectiveness, emotional safety, and ethical and data privacy concerns. We ask: \emph{How do Chinese graduate students evaluate AI-CBT chatbots as self-help tools, and which factors shape their intentions to use (or avoid) them?}

\paragraph{Theoretical Framing.}
We draw on two complementary models. The Health Belief Model (HBM) posits that health actions reflect perceived susceptibility, perceived severity, perceived benefits and barriers, signals to action, and self-efficacy \cite{Becker1974,Carpenter2010}. In the context of student well-being, HBM helps explain how learners assess everyday emotional challenges (e.g. stress, low mood), balance anticipated benefits of AI-CBT (e.g., immediacy, privacy) against barriers (e.g., data security, depersonalization), and identify triggers that prompt trial or sustained use. HBM has been applied to mental health-relevant behaviors, including exercise promotion \cite{csahan2024effect}, pandemic coping \cite{Zhou2021}, and belief formation \cite{kroke2022conspiracy}. The Theory of Planned Behavior (TPB) complements HBM by modelling behavioural intention as a function of attitudes toward the behaviour, subjective norms, and perceived behavioural control \cite{Ajzen1991,Ajzen2020}. In this study, TPB clarifies how favorable or unfavorable evaluations of AI-CBT (usefulness, credibility, emotional safety), perceived social expectations (peers, supervisors, family) and perceived control (confidence in privately selecting and using a chatbot) converge on intentions to adopt or avoid such tools. Together, HBM and TPB offer a structured lens for interpreting psychological, social, and contextual determinants of AI-CBT acceptance in a nonclinical high-pressure academic setting.

\paragraph{Context and Gaps.}
Evidence in Chinese university students shows that stigma negatively predicts help seeking, while mental health literacy and social support exert positive effects (directly and through chain mediation) \cite{yang2024relationship}. Students often equate counseling with 'serious problems' and report concerns about confidentiality, professionalism, and power dynamics in campus services \cite{ning2024chinese}. In parallel, some Chinese and Chinese American students report turning to anonymous, technology-mediated supports for privacy and immediacy, while expressing caution about generic responses from automated systems \cite{zhang2025use}. These patterns suggest that trust, perceived effectiveness, emotional safety, and data or ethical assurances are likely to be central to adoption decisions, but little is known about how these considerations are weighed in the daily realities of graduate study in China. In addition, survey-based studies with general or clinical samples \cite{Varghese2024,Benda2024} cannot easily capture the meanings, contingencies, and trade-offs that shape students' decisions in practice.

\paragraph{Approach}
To address these gaps, we adopt a qualitative interpretive design using semi-structured interviews with Chinese graduate students who have experience with, or exposure to, AI-driven mental health tools. Reflexive thematic analysis is used to reveal patterns in how participants construct trust and usefulness, negotiate emotional and ethical boundaries, and interpret social expectations and control in daily use. HBM and TPB serve as sensitizing frameworks during interpretation, supporting a theoretically informed synthesis of perceived benefits or barriers, cues to action, attitudes, norms, and perceived control as they relate to the adoption of AI-CBT.

\paragraph{Contributions.}
This paper makes three contributions:

\begin{itemize}
\item \textbf{Empirical:} It provides a focused account of Chinese graduate students’ perceptions of AI-CBT chatbots, an underrepresented, non-clinical, high-stress cohort, grounded in first-person experiences rather than inferred from general population surveys.
\item \textbf{Conceptual:} It integrates HBM and TPB to organize and interpret qualitative accounts of AI-CBT evaluation, clarifying how perceived benefits/barriers, cues to action, attitudes, norms, and perceived control co-produce intentions in this context.
\item \textbf{Practical:} It distils design-relevant implications, in particular the need for transparent data handling assurances, safeguards for emotional safety, culturally tuned onboarding, and clear pathways for escalation, tailored to student users in China and comparable settings.
\end{itemize}


\section{Literature Review}

\subsection{Public perceptions of AI in mental health}
Mental health tools enabled by AI are increasingly being positioned as scalable and always available complements to traditional care, with potential benefits in accessibility, cost-effectiveness, and stigma reduction \cite{Benda2024,Varghese2024,Graham2019,Higgins2023,Torous2014}. At the same time, there are constant concerns about data privacy, precision, explainability, and perceived loss of `human touch' \cite{Benda2024,Chew2021,Varghese2024,Gerlich2023,Wu2023}. Surveys and syntheses suggest that acceptance varies between sociodemographic groups and with prior exposure to digital technologies \cite{Benda2024,Beets2023}; health professionals similarly express guarded optimism tempered by ethical and safety reservations \cite{Rogan2024,Fiske2018}. In general, awareness in the general public is still uneven, but there is a growing openness to using AI-supported tools for prevention, self-monitoring, stress management, and everyday maintenance of wellness \cite{Gerlich2023,Varghese2024,Wu2023}. These patterns highlight a familiar trade-off: perceived utility and convenience on the one hand, set against trust, safety, and ethical assurance on the other.

\subsection{AI–CBT chatbots: efficacy, engagement, and acceptability}
Within digital mental health, AI-powered CBT (AI–CBT) chatbots have shown measurable benefits for symptom reduction and engagement in trials and real-world deployments \cite{Boucher2021,He2022,Inkster2018}. Recent evaluations report clinically significant effects, for example, reductions in depressive symptoms shortly after use with sustained improvement at follow-up, and significant reductions in anxiety during guided chatbot sessions \cite{He2022,mcfadyen2024ai}. In some comparisons, AI-supported tools have achieved higher rates of reliable improvement and recovery than standard CBT pathways \cite{Habicht2025}. Beyond outcomes, qualitative and mixed methods studies indicate that users often value immediacy, structure, and self-paced support \cite{Inkster2018,Boucher2021}. However, privacy and governance concerns remain salient barriers \cite{Sun2019}, and `trust in automation' continues to be a central determinant of intention and sustained engagement \cite{Jian2000}. Reviews also note heterogeneity in study quality, outcomes and reporting standards, underscoring the need for context-specific evidence and a stronger theoretical basis \cite{Boucher2021,ZhiChen2024,ZhangWang2024,Chan2021,Patel2019}.

\subsection{Limitations of the current evidence base}
Much of the published work on AI-driven mental wellness tools relies on self-reported cross-sectional survey data, which can be affected by response biases even when anonymity helps mitigate social desirability \cite{Beets2023,Benda2024}. A substantial portion of the literature focuses on perceived effectiveness and trust in clinical or broad public samples, providing limited information on nonclinical, high-pressure populations such as graduate students \cite{ApolinarioHagen2017,Boucher2021,Varghese2024}. Many syntheses also draw on studies conducted prior to the latest wave of generative AI advances, raising questions about temporal relevance \cite{Aggarwal2020,Beets2023,Gerlich2023,Wu2023}. In evaluations that report high acceptability and satisfaction, there is a plausible survivor (self-selection) bias because feedback often comes from those who already chose to use the apps; perspectives of non-users and those deterred by trust, privacy, or cultural factors are underrepresented \cite{Habicht2025,Inkster2018,Boucher2021}. Privacy and data stewardship remain incompletely addressed, despite their centrality to adoption decisions \cite{Sun2019}. Collectively, these limitations point to the need for qualitative, theoretically informed work that can unpack how users interpret trust, effectiveness, and emotional safety in real-world decisions about AI–CBT, especially in non-Western contexts.

\subsection{Positioning and contribution}
Although recent surveys document broad attitudes towards AI in mental health and identify demographic correlates of acceptance \cite{Benda2024,Varghese2024}, much less is known about how psychological determinants shape intentions to use AI–CBT in self-help settings. Health behavior theories such as the Health Belief Model (HBM) and the Theory of Planned Behavior (TPB) offer well-validated constructs: perceived susceptibility, severity, benefits, barriers, self-efficacy; and attitudes, subjective norms, perceived control that can explain adoption decisions \cite{Becker1974,Carpenter2010,Ajzen1991,Ajzen2020}. However, these frameworks are most often applied through quantitative surveys, leaving a gap in qualitative accounts that illuminate how users make sense of these constructs in everyday contexts. Addressing this gap, the present study uses semi-structured interviews, interpreted through HBM and TPB, to examine how Chinese graduate students in academically demanding environments perceive and engage with AI–CBT chatbots. Using trust, perceived effectiveness, emotional safety, data privacy, and social norms, the study aims to provide culturally and contextually grounded insights that can inform the design, communication, and implementation of AI-based mental well-being tools for student populations in China.

\section{Methods}\label{sec3}

\subsection{Design and rationale}
This study used a qualitative design with semistructured one-to-one interviews to examine how Chinese graduate students perceive and experience AI-powered cognitive behavioral therapy (AI-CBT) chatbots for mental well-being. The approach was chosen to capture situated meanings, decision processes, and contextual nuances that fixed response surveys cannot easily reveal. The analysis followed Braun and Clarke’s reflexive thematic analysis \cite{BraunClarke2021}, allowing codes and themes to develop from the data while using the Health Belief Model (HBM) and the Theory of Planned Behavior (TPB) as sensitizing frameworks \cite{Becker1974,Carpenter2010,Ajzen1991,Ajzen2020}. These frameworks guided the interpretation of perceived benefits and barriers, self-efficacy, attitudes, social influence, perceived control, and intentions.

\subsection{Participants and recruitment}
Ten postgraduate students of Chinese ethnicity, enrolled in master's or doctoral programs in mainland China, Hong Kong, or Macau, were recruited through non-incentivized convenience sampling through WeChat and student networks. Inclusion required being at least 18 years old, fluent in Chinese or English, and familiar with AI-enabled applications such as chatbots, virtual assistants, or digital mental wellness tools. Individuals were excluded if they were under 18 years old, fluent in Chinese or English, had no prior exposure to AI tools, or self-reported a current or previous clinical diagnosis of a mental health disorder. These criteria ensured that participants could engage meaningfully with the topic, maintained a nonclinical focus, and reduced the risk of distress. A target of ten interviews was set a priori as appropriate for exploratory thematic work, suitability being judged by thematic sufficiency rather than statistical generalizability \cite{Guest2006}. All invited participants completed the study and there was no attrition.

\subsection{Procedure and data collection}
Ethical approval was obtained prior to field work (HSEARS20250916001). Interested individuals received an online information sheet and provided electronic consent. The interviews were conducted on in-person, recorded audio with permission, and lasted about 15 to 30 minutes. The semi-structured guide, informed by HBM and TPB, invited reflection on perceived usefulness, anticipated benefits and barriers, trust and emotional safety, data privacy and ethical concerns, social expectations, perceived control, and intentions to use AI-CBT. The questions were open-ended to encourage narrative depth. The interviewer adopted an empathetic, non-directive stance to support comfort and disclosure. No procedural changes were required during data collection. 

\subsection{Transcription and data management}
The recordings were transcribed verbatim using \textit{iflyrec} (v25.8.2250) and then manually checked for accuracy. The transcripts were anonymized by replacing names and removing direct identifiers. All files were stored on encrypted drives with restricted access to the researcher. Units of analysis consisted of narrative segments rich in meaning within each transcript so that context and interpretive depth were preserved.

\subsection{Analysis}
Reflexive thematic analysis proceeded through six phases: (1) familiarization through repeated reading and memo writing; (2) initial coding close to the data; (3) construction of candidate themes by collating related codes; (4) iterative review of coherence within themes and distinction between themes; (5) definition and naming with clear boundaries and scope; and (6) production of an analytic narrative supported by illustrative extracts, following Braun and Clarke’s guidance \cite{BraunClarke2021}. Coding started inductively. After the themes stabilized, a secondary interpretive pass used HBM and TPB as sensitizing concepts to clarify how the accounts of the participants corresponded to perceived benefits and barriers, self-efficacy, attitudes, subjective norms, perceived behavioral control and intentions \cite{Becker1974,Carpenter2010,Ajzen1991,Ajzen2020}. A simple matrix linked themes and subthemes to these constructs, making the theory–data connections explicit.

\subsection{Methodological integrity}
Rigor was supported by an audit trail that documented coding decisions, evolving theme maps, and analytic memos. The researcher maintained a reflexive journal to surface assumptions and considered their influence on interpretation and engaged in periodic peer debriefing with a supervisor to investigate alternative readings and test the stability of claims. Earlier transcripts were revisited as the themes evolved to maintain analytic consistency. In line with reflexive thematic analysis, the credibility was based on transparency, thick description, and coherent argumentation rather than coefficients between parties \cite{BraunClarke2021}. Data collection was considered sufficient when additional interviews no longer yielded new codes or challenged the thematic structure, indicating thematic sufficiency for the study objectives \cite{Guest2006}.

\section{Thematic Findings and Results}
\subsection{Thematic Findings}
\subsubsection*{Theme 1: ``It’s There When I Need It'' — Perceived Usefulness and Accessibility}

Participants emphasized the convenience and immediacy of AI-CBT chatbots, often highlighting their availability 24 hours a day and freedom from scheduling or stigma-related barriers. Chatbots were described as a low-cost, low-stigma resource that could be accessed at any time, particularly when students felt low late at night or during stressful deadlines. For many, AI-CBT provided a space for minor emotional release and situational support rather than intensive therapy. As one participant explained that ``Sometimes I just need to vent a little before sleeping... it is nice that I can talk to the chatbot anytime'' (P3). Another noted ``I don’t need to make an appointment or wait; it’s just there when I need it'' (P7). This framing situates AI-CBT as a practical coping aid for everyday stressors, though with bounded utility for deeper concerns.

\subsubsection*{Theme 2: ``It’s Not a Real Person'' — Limitations of Empathy and Trust}

Although anonymity and lack of judgment were valued, participants consistently questioned the ability of the chatbot to provide emotional depth, empathy, and personalized support. The students described the responses as overly rational, repetitive, or generic, which undermined trust in the effectiveness of the tool for complex issues. One participant mentioned that ``He may not give too much emotional value… every time I ask, he gives me the same kind of advice'' (P1). Another participant said that ``It listens, but sometimes I feel like I am just typing into a machin'' (P5). The lack of a ``human touch'' was a recurring limitation, leaving participants uncertain about the long-term therapeutic value of the chatbot.

\subsubsection*{Theme 3: ``Is My Data Safe?'' — Privacy and Ethical Concerns}

Concerns about data security and confidentiality emerged as a strong deterrent to adoption. Many participants reported self-censoring sensitive disclosures due to uncertainty about how data could be used by institutions or companies. As one of the participants explained that ``“I don’t know where the data go, so I don’t say anything too private'' (P1). Another participant worried about institutional oversight, ``If it’s just to help me, fine. But what if it is used by companies or schools?'' (P4). Such privacy concerns directly limited the perceived usefulness of AI-CBT and weakened the willingness of students to rely on the tool for meaningful engagement.

\subsubsection*{Theme 4: ``It’s for Others, Not for Me'' — Social Norms and Cultural Stigma}

Adoption was shaped by cultural stigma and peer expectations. Some participants positioned AI-CBT as more suitable for 'others' who were lonely or introverted, distancing themselves from direct use. This reflected the ongoing stigma surrounding mental health support. As a participant said: ``It’s good for people who don’t want to talk to others, but I still feel strange using it'' (P6). Concerns about social reputation also surfaced: ``If my classmates knew that I used it, they might think I have problems'' (P8). Although some family members and supervisors were seen as neutral or even supportive, the anticipated peer judgment constrained adoption.

\subsubsection*{Theme 4: ``I'd Try it, But Only If...'' — Conditions for Adoption and Self-efficacy}
The students varied in their confidence (self-efficacy) in the use of AI-CBT. Some expressed high readiness, while others were hesitant due to doubts about effectiveness or unfamiliarity with such tools. Importantly, the participants outlined clear conditions for adoption: transparent privacy safeguards, more natural and empathetic interactions, and culturally relevant design. The participants said that ``If it can guarantee that my information will not be leaked, then I would feel okay'' (P2) and ``If the system felt more human and easier to use, I’d be more willing'' (P4).For many, AI-CBT was positioned as a complementary tool rather than a replacement for professional therapy, useful for managing surface-level issues, but not suited for deeper psychological needs.

\section{Discussion}

This study explored how Chinese postgraduate students perceive and evaluate AI-powered cognitive behavioral therapy (AI-CBT) chatbots for mental wellbeing. The analysis identified five interrelated themes: (1) \textit{``It’s There When I Need It''} — usefulness and accessibility; (2) \textit{``It’s Not a Real Person''} — limits of empathy and trust; (3) \textit{``Is My Data Safe?''} — privacy and ethical concerns; (4) \textit{``It’s for Others, Not for Me''} — social norms and cultural stigma; and (5) \textit{``I’d Try It, But Only If…''} — conditions for adoption and self-efficacy. Taken together, these themes highlight a cautious openness to AI-CBT, tempered by concerns over privacy, trust, and cultural stigma.

\subsection{Interpreting the Key Findings through HBM and TPB}

\subsubsection{AI-CBT as a Low-Stigma, Accessible Resource:}
Students consistently valued the immediacy and 24/7 availability of AI-CBT, which aligned with perceived benefits in HBM and favorable attitudes in TPB. Convenience and anonymity were described as particularly important when traditional help seeker felt too costly, stigmatizing, or inaccessible. Similarly to previous findings, participants placed AI-CBT as a supplementary low barrier coping aid to manage everyday stress. However, this perceived utility was bounded: the tool was framed as suitable for 'light' issues but not for addressing deeper psychological concerns.


\subsubsection{Trust, Empathy, and Emotional Safety as Barriers:}
A recurrent limitation was the perception that AI-CBT lacks empathy and emotional attunement. Students described chatbot responses as repetitive, generic, and overly rational, which undermined trust and reduced confidence in achieving meaningful support. Within the HBM, this represents a barrier that diminishes self-efficacy; in the TPB, it weakens perceived behavioral control. Previous studies similarly note that the absence of human warmth can lead to skepticism about the therapeutic potential of AI. This suggests that for AI-CBT to be effective, interaction design must better approximate empathic understanding and personalized guidance.


\subsubsection{Privacy and Ethical Concerns as Adoption Constraints:}
Data security was a dominant theme, and many participants explicitly withheld sensitive disclosures due to uncertainty about data governance. This aligns with the HBM concept of barriers, directly constraining adoption intentions. In TPB terms, privacy concerns negatively shaped attitudes and intentions to use AI-CBT. This finding echoes broader evidence that privacy and transparency are central determinants of digital mental health adoption. Unless addressed through clear, accessible safeguards, data concerns can prevent students from engaging sufficiently for therapeutic benefit.


\subsubsection{Stigma, Social Norms, and the Role of Cultural Expectations:}
The accounts of the participants reflect the ongoing cultural stigma surrounding help seeking. Although some family members and supervisors were perceived as neutral, anticipated peer judgment was a strong deterrent. In terms of TPB, this reflects subjective norms that weaken intentions despite positive attitudes. Similarly to previous research, students often framed AI-CBT as something for 'others' in need, distancing themselves from direct use. These findings underscore the importance of normalizing digital mental health tools within academic and cultural contexts, to reduce perceived stigma and reputational risks.

\subsubsection{Conditions for Adoption and the Boundaries of Self-Efficacy:}
Confidence in effectively using AI-CBT varied. Some participants expressed a high readiness, while others doubted its usefulness or felt uncertain about how to participate. This reflects variability in self-efficacy (HBM) and perceived behavioral control (TPB). Importantly, the participants articulated specific conditions for adoption: trustworthy privacy safeguards, culturally adapted content, and more human-like interaction. These findings parallel previous evidence that students' willingness to participate is contingent on design characteristics that promote trust and perceived fit.

\subsection{Integrating HBM and TPB: Explaining Conditional Openness}
The combined use of HBM and TPB clarifies how students’ adoption decisions are shaped. HBM explains the balance of perceived benefits (accessibility, privacy, immediacy) against barriers (lack of empathy, data concerns, stigma), alongside signals for action such as curiosity, loneliness, or peer encouragement. TPB further illuminates how positive attitudes are constrained by stigmatized norms and reduced perceived control due to trust and empathy limitations. Together, these models highlight why students express conditional openness: willing to try AI-CBT in principle, but hesitant to rely on it without safeguards, normalization, and stronger assurances of trust.

\subsection{Implications for Design, Deployment, and Policy}

The implications of the findings can be categorized into five focus areas: guidance design, privacy, normalization, scope / transfer and cultural fit, each paired with concrete actions to improve trust, usability, and ethical implementation of AI-CBT in university contexts, as shown in Table 1.

\begin{table}[ht!]
\centering
\renewcommand{\arraystretch}{1.0}
\small 
\begin{tabular}{p{3.5cm} p{8.5cm}}
\hline
\textbf{Focus} & \textbf{Actionable Guidance} \\
\hline
\textbf{Design for attuned guidance (trust \& efficacy)} & 
Add context-aware scaffolding (brief reflective summaries, selectable response styles, culturally relevant examples) to increase felt understanding and self-efficacy. Provide in-flow transparency cues (``why this question,'' ``how this suggestion was generated'') to strengthen epistemic trust without disrupting conversation. \\ \hline

\textbf{Make privacy legible and actionable (reduce barriers)} &
Convert privacy policies into just-in-time, plain-language micro-disclosures (e.g., ``Stored only on your device unless you opt in''). Offer concrete controls (local-only mode, one-tap deletion, redaction sliders) and visible data-action logs so users can manage risk, not guess about it. \\ \hline

\textbf{Normalize use within academic ecosystems (shift norms)} &
Integrate AI-CBT alongside study/wellness resources in university portals as a skills tool—not a clinical substitute. Run opt-in peer-mentor/counselor campaigns that highlight everyday, low-stakes scenarios (exam stress, sleep hygiene) to reframe norms from ``help-seeking = weakness'' to ``self-management = professionalism.'' \\ \hline

\textbf{Clarify scope and handoffs (safe boundaries)} &
Communicate scope clearly (supports mild–moderate distress; not crisis care). Build visible escalation paths to human services when risk signals emerge (proactive prompts, one-click referral), preserving emotional safety and safeguarding users. \\ \hline

\textbf{Culturally responsive content (fit to context)} &
Localize metaphors, examples, and coping strategies to Chinese graduate-school realities (advisor dynamics, lab deadlines, candidacy exams). Emphasize relevance to increase perceived benefits and improve attitudes toward use. \\
\hline
\end{tabular}
\caption{Two-column summary of implications for AI-CBT design, deployment, and policy in Chinese graduate-student contexts.}
\end{table}

\section{Conclusion}
This study shows that five interrelated themes shape the way Chinese post-graduates consider using AI-CBT. From an HBM perspective, increases in stress or time pressure act as cues for action, while the benefits of immediacy, constant availability, and reduced interpersonal exposure encourage acceptance. At the same time, barriers such as concerns about privacy, data use, and limited empathy restrict participation. From a TPB perspective, trust and emotional safety improve the sense of control of students and strengthen their intentions, whereas stigma and negative social norms can override positive attitudes and discourage use.

Taken together, the findings suggest that sustained engagement with AI-CBT will depend on three factors: (i) building trust through transparent data practices and explainable responses, (ii) strengthening perceived control through clear guidance, supportive interaction and safety assurances, and (iii) normalizing use within academic communities to reduce stigma.

Empirically, this research provides a context-specific account of how Chinese graduate students weigh convenience against emotional, privacy, and cultural constraints. Theoretically, it shows how the benefits-barrier trade-offs of HBM intersect with TPB's attitudes, norms, and perceived control under conditions of cultural stigma. Methodologically, the qualitative approach adds depth to predominantly survey-based studies of digital mental health acceptance, highlighting practical leverage points for both design (e.g., empathic interaction, privacy-centered UX) and implementation (e.g., campus-level signposting and norm setting) in high-pressure educational contexts.


\paragraph{\textbf{Limitations and future work.}} The findings arise from a small, nonprobability sample of Chinese graduate students and self-reported interview data; they are not intended to generalize statistically.  Future studies should (a) test these pathway hypotheses with larger and more diverse samples, including longitudinal or mixed-method designs; (b) examine cross-cultural cohorts to disentangle cultural norms from technology perceptions; (c) experimentally evaluate design characteristics (e.g., anthropomorphism, tone, privacy disclosures) on trust, emotional safety, and intention; and (d) co-design and trial institutionally supported AI-CBT deployments that pair transparent data practices with stigma-reducing communications, tracking both engagement and wellbeing outcomes over time.

%
%
%

\end{document}